\documentstyle[aps,prl,epsfig,multicol]{revtex}

\newif\ifboo \boofalse

\begin{document}

\title{Identification of clusters of companies in stock indices via Potts 
super-paramagnetic transitions}

\author{L. Kullmann$^1$, J. Kert\' esz$^{1,2}$, R. N. Mantegna$^3$}
\address{$^1$Department of Theoretical Physics, Institute of Physics, 
Technical University of Budapest, 8 Budafoki út, H-1111 Budapest, Hungary}
\address{$^2$Laboratory of Computational Engineering, Helsinki University of Technology, 
P.O.Box 9400, FIN-02015, Espoo, Helsinki, Finland}
\address{$^3$Istituto Nazionale per la Fisica della Materia, Unit\`a di Palermo
and Dipartimento di Energetica ed Applicazioni di Fisica, Universit\`a di Palermo, 
Palermo, I-90128, Italy}
\date{\today}

\maketitle

\begin{abstract}
The clustering of companies within a specific stock market index is
studied by means of super-paramagnetic transitions of an
appropriate $q$-state Potts model where the spins correspond to
companies and the interactions are functions of the
correlation coefficients determined from the time dependence of the
companies' individual stock prices. The method is a generalization of
the clustering algorithm by Domany {\it et. al.} to the case of
anti-ferromagnetic interactions corresponding to
anti-correlations. For the Dow Jones Industrial Average where no
anti-correlations were observed in the investigated time period, the
previous results obtained by different tools were well reproduced. For
the Standard \& Poor's 500, where anti-correlations occur, repulsion
between stocks modify the cluster structure.
\end{abstract}

\begin{multicols}{2}

Stock market indices, like the Dow Jones (DJ) or Standard \& Poor's
500 (S\&P 500) are used as indicators of the status of the
markets. They are averaged values of a different number of selected
companies indicative of the economy of a given market. It is of both
theoretical and practical importance to analyze how individual
contributions to the average behave. The customary approach in the
financial literature focuses on the investigation of the properties of
the covariance matrix. Here we take a different approach aiming to
identify the presence of a hierarchical structure inside the set of
stocks simultaneously traded in a market. The identification of the
hierarchy of clusters is of central importance both from the point of
view of understanding the dynamics of the stock index and for
portfolio optimization~\cite{EG,BP}. As far as we know this question
was first analyzed by Mantegna by means of the minimal spanning tree
method~\cite{M1,M2,MS}, see also~\cite{BVM}.  Here we analyze the
problem of clustering of companies in the S\&P 500 and the DJ indices
by a different method based on the q-state Potts model which turns out
to be particularly suitable to handle anti-correlations.

The idea to use the super-paramagnetic (SPM) ordering of a q-state
Potts model for cluster identification is due to Domany {\it
et. al}~\cite{BWD,WBD,D}. They start from a set of points which lie in a
metric space where the mutual distances of the points are known. By
introducing a distance dependent ferromagnetic (FM) interaction
between Potts spins assigned to the points at appropriately chosen
temperatures the close points within a cluster feel strong interaction
and align while far clusters point into different "Potts
directions". The functional dependence of the interaction on the
distance should be chosen in an appropriate way. For a given
interaction the possible hierarchic clustering shows up in a series of
SPM transitions.

We have generalized this method by dropping the condition of the
metric and allowing negative (anti-ferromagnetic, AFM) couplings. The
coupling between the pair of Potts spins ({\it i.e.} companies) is in our
case the explicit function of the correlation coefficient and it is FM
for positive correlations and AFM for anti-correlations
(the latter are present in the S\&P 500).  
This way we naturally take into account the "repulsion" between
negatively correlated companies (and clusters of companies) -- an
important aspect for portfolio optimization. In order to estimate the
effect of the anti-correlations we carried out calculations with only
FM interactions ({\it i.e.} we took the absolute values of the correlation
coefficients) and with correctly signed interactions too. We found
that the difference -- in our set of data -- can be observed only at
the ground state,
{\it i.e.} for the main (dominant) cluster structure.

Consider a $q$ state, inhomogeneous Potts model: $s_i = 1, ..., q$,
where $i = 1, ..., N$. $N$ is the number of points one should arrange
in clusters (the number of companies in the considered situation). The cost
function will be the Hamiltonian:
\begin{equation}
\label{hamiltoni}
H = - \sum_{(i,j)} J_{ij} \delta_{s_i,s_j}.  
\end{equation}

The coupling $J_{ij}$ is a function of the correlation coefficient
$c_{ij}$ between the time evolution of the logarithmic daily
price return $Y_i = \log(P_i(t)) - \log(P_i(t-1))$ of the stock of
companies $i$ and $j$. $P_i(t)$ is the closure price of the stock $i$ at
the day $t$. The correlation coefficient can be computed as
follows:
\begin{equation}
c_{ij} = { \langle Y_i Y_j \rangle - \langle Y_i \rangle\langle Y_j \rangle
\over \sqrt{ 
\left( \langle Y_i^2 \rangle - \langle Y_i \rangle^2 \right)
\left( \langle Y_j^2 \rangle - \langle Y_j \rangle^2 \right)
}
} \;\; \in [-1,1].
\end{equation}
Here $\langle ... \rangle$ is a temporal average performed on all the
trading days of the investigated time period which ranges from July
3rd, 1989 to October 27th, 1995.

The Potts model can be used for cluster identification in the
following way. Let us first consider the simpler, FM case \footnote{It
should be mentioned here that there are no anti-correlations in the DJ
index in the investigated time period.}, {\it i.e.}, $J_{ij} \geq 0$.
These couplings are functions of the property the clustering should be
based upon -- in our case this is the correlation coefficient.  If the
set of spins are interrelated in a way that each pair of spins can be
connected through a path via non-vanishing $J_{ij}$-s the ground state
of the system is all spins pointing into one Potts-direction, {\it
i.e.}, they build a single cluster.  As the temperature is increased,
weak bonds break easier than the strong ones and transition to a SPM
phase takes place where clusters of spins have a specific Potts
magnetization but the net magnetization of the whole system is
zero. The clusters identified in this manner are those we have been
looking for. Depending on the interactions, the system may go through
a sequence of such transitions signalizing the hierarchical cluster
structure.  The transitions are best indicated by monitoring the peak
structure in the susceptibility and the clusters are then identified
by means of the spin-spin correlation functions.

The method is easily generalized to the case where repulsion between
pairs of points is present, in our case, if there are anti-correlations
between companies as it is the case {\it e.g.} for the S\&P 500. This
latter case is important when AFM interactions are also
present, because of low temperature behaviour.

For the above reasons we make the following choice for the interaction:
\begin{equation}
\label{interaction}
J_{ij} = \mbox{sgn}(c_{ij}) \left(
1 - \exp \left\{ - {n-1 \over n} \left[{c_{ij} \over a}\right]^n \right\}
\right),
\end{equation}
where $c_{ij}$ is the correlation coefficient between companies $i$
and $j$.
The parameters ($a$, $n$) should be chosen so
that the super-paramagnetic state exist, but inside this region the
result will be not too sensible on the choice. The fine tuning serves
to be able to observe the transitions more clearly, {\rm i.e.} make peaks 
in the susceptibility function sharper, and the constant regions between them
larger.
A possible determination of parameter $a$ is the average of the largest
correlation coefficients for each spin: $a=1/N \sum_{i=1}^N
\max_j (c_{ij})$. The power $n$ tunes the range of interaction, the
factor $n/(n-1)$ in the exponent shifts the inflection point of the
interaction function. 

The order parameter is:
\begin{equation}
\label{order_param}
m = {N_{max}/N - 1/q \over 1 - 1/q},
\end{equation}
where $q$ is the number of states that a spin can have, $N$ is the
number of spins, and $N_{max}$ is the maximal number of spins which
are in the same state. The value of the parameter $q$ is
determined by a trial and error optimization: Too small $q$ hinders
the identification of the SPM clusters since different
clusters are then forced to point into the same ``Potts
direction''. Too large $q$ makes the calculations more cumbersome. The
results depend only weakly on the value of $q$.

The (first order-like) SPM transitions are identified
by the peaks in the susceptibility $\chi = N(\langle
m^2\rangle - \langle m \rangle^2 )/T$, which has to be measured as the
function of the temperature. (For more convenience we studied the 
$\hat{\chi}(T)=T\chi(T)/N$ function.)
In the simplest non-trivial case there are two
transitions as temperature is increased: FM $\to$ SPM and SPM $\to$ PM.
However, inside the range of the
SPM phase one can often observe several peaks too, meaning that
there are more than one characteristic cluster
configurations. As the
temperature increases the clusters 
break up into sub-clusters: A hierarchical structure is revealed.

The susceptibility $\hat{\chi}$, is approximately constant between two peaks. At
this temperature regime the clusters are identified by means of the
spin-spin correlation function: If the correlations between two spins
exceed a given threshold ({\it e.g.} $0.5$), the spins (companies) are
considered to belong to the same cluster. The result is not sensitive
on the choice of the threshold value. The distribution of the
spin-spin correlation has two peaks, one of them is near to the value
zero and the other near to the value one. The probability that the
correlation of two spins lies between these two values is low.

\begin{figure}
\centerline{
\epsfig{file=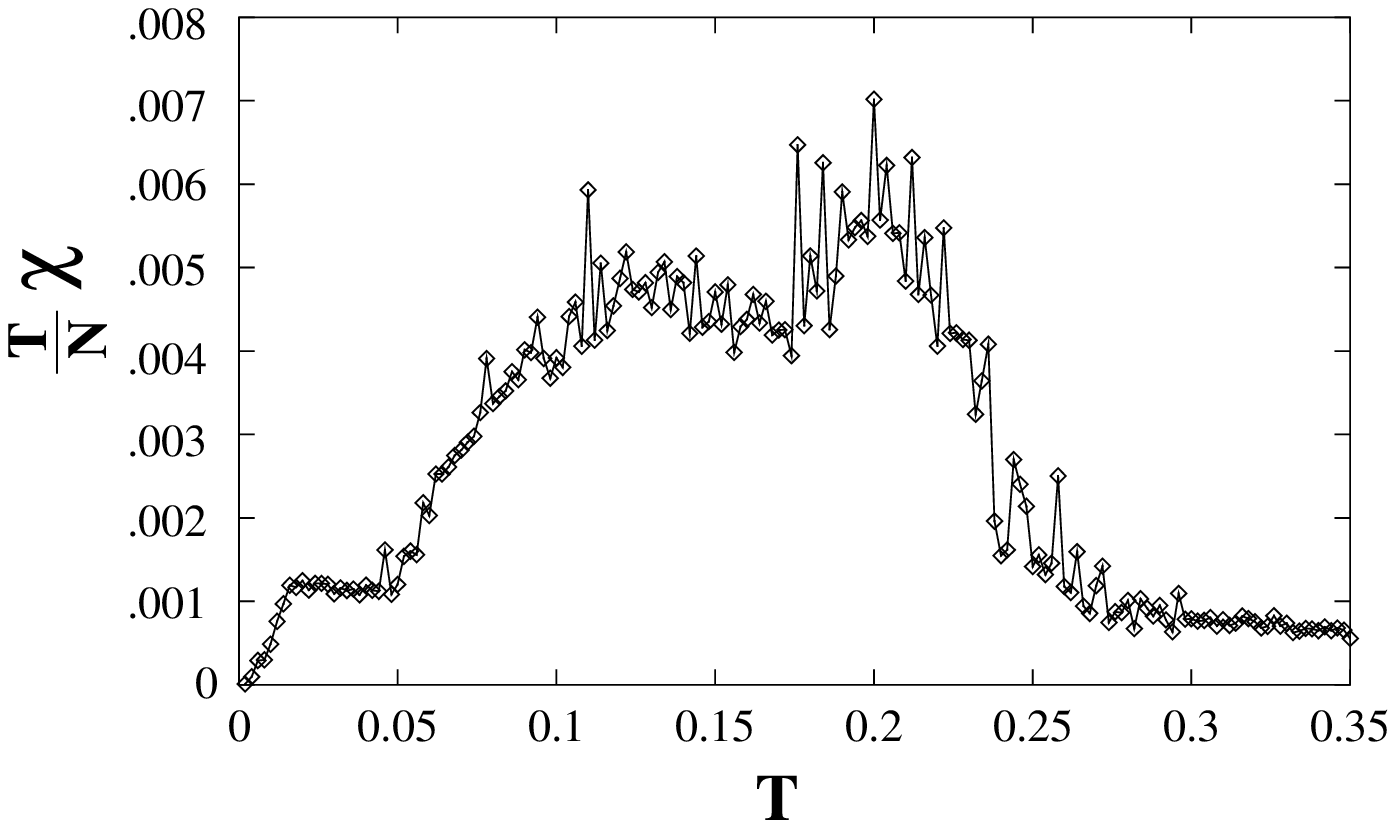,height=1.8truein,width=3.0truein}}
\caption{\footnotesize Temperature dependence of $\hat{\chi}=T\chi/N$ 
in the FM case
(Dow Jones companies). The parameters are: $a=0.43,\; n=8$ and $q=100$. }
\label{fm_dj_khi}
\end{figure}

First we analyzed the companies of the Dow Jones index which includes
$N=30$ companies.
No negative correlation coefficient was found for this
data set ($c_{ij}\geq 0$). In this purely FM case the simulation
could be done with the efficient Swendsen-Wang method~\cite{SW}.
The parameters of the interaction (\ref{interaction}) were set to
$a=0.43,\; n=8$.
The temperature
dependence of the susceptibility $\hat{\chi}$, is shown in 
Fig.~\ref{fm_dj_khi}.

We analyzed the spin-spin correlations at four different temperature levels.
We got the following clusters (we use the standard notation for the companies):\\
{
\small
$T=0.03$\\
1) AA, ALD, AXP, BA, BS, CAT, CHV, DD, DIS, GE, GM, IP, JPM, KO, MCD,
MMM, MO, MRK, PG, T, TX, UK, UTX, XON,\\
$T=0.13$\\
1) AA, DD, GE, GM, IP, JPM, KO, MCD, MMM, MO, MRK, PG, T \\
2) CHV, TX, XON\\
$T=0.15$\\
1) CHV, TX, XON\\
2) DD, GE, IP, KO, MCD, MMM, MO, MRK, PG, T\\
$T=0.28$\\
1) CHV, TX, XON
}

Due to the small system size we have large fluctuations and the
identification of clusters is not unambiguous. Nevertheless, the
clusters do well match with the deterministic minimal spanning tree
calculation by Mantegna though the hierarchical structure of the
clusters is not so clearly represented.

\begin{figure}
\centerline{
\epsfig{file=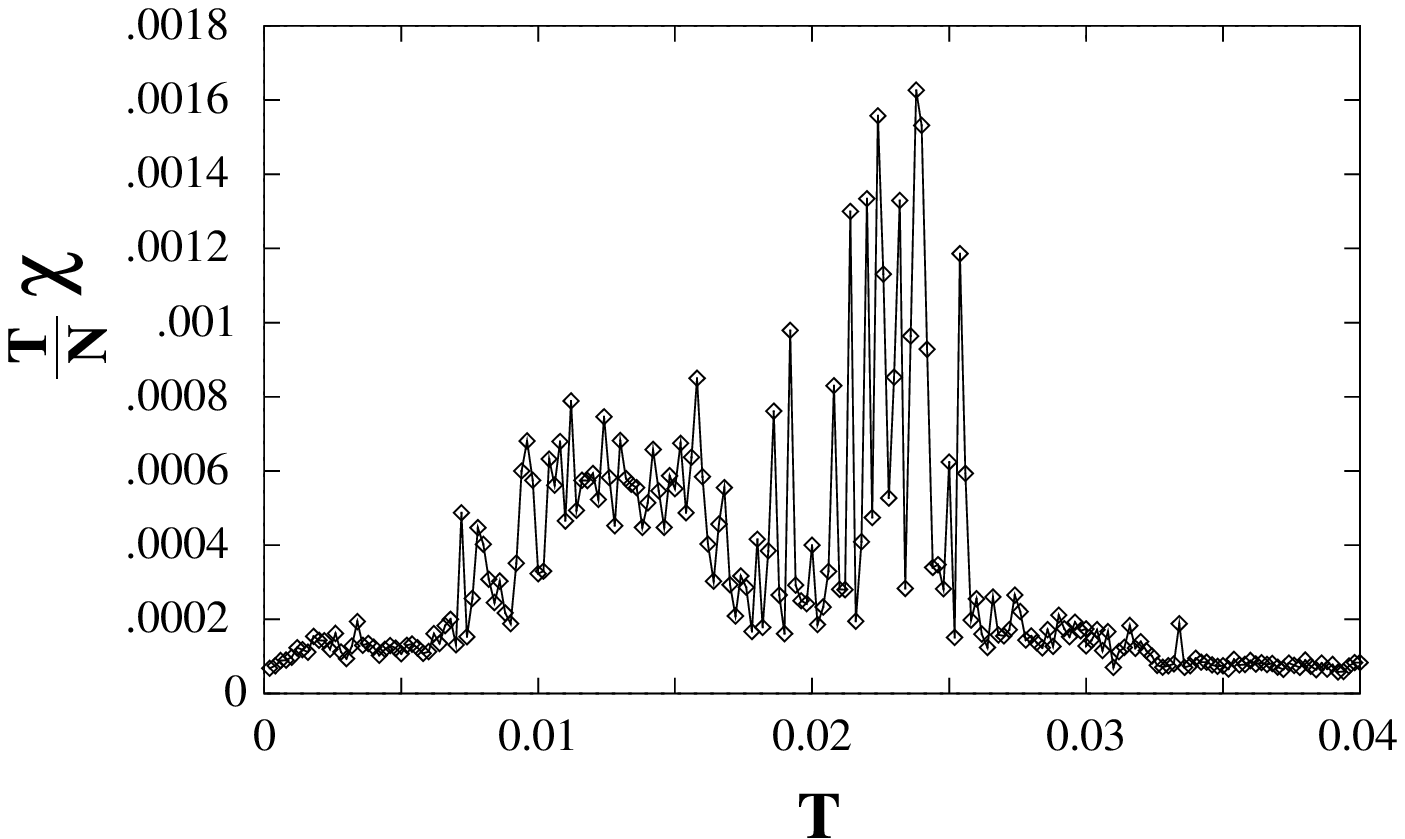,height=2.0truein,width=3.0truein}}
\centerline{}
\caption{\footnotesize Temperature dependence of $\hat{\chi}=T\chi/N$ 
in the FM case (S\&P500 companies)
with the parameters: $a=0.65,\; n=10$ and $q=50$.}
\label{fm_sp_khi}
\end{figure}

\begin{figure}
\centerline{
\epsfig{file=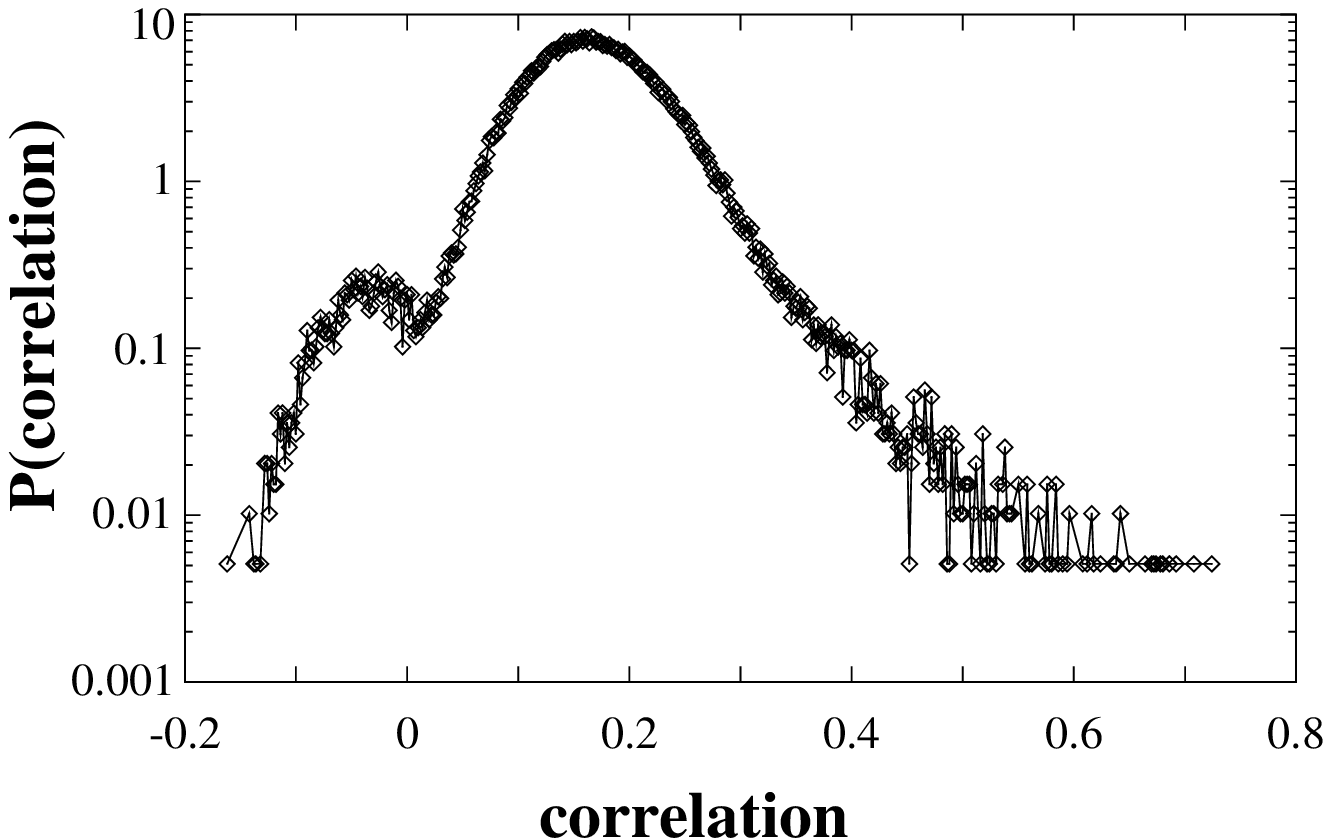,height=2.0truein,width=3.0truein}}
\centerline{}
\caption{\footnotesize Distribution of the measured correlation coefficients 
(S\&P500 companies).}
\label{corr_distrib}
\end{figure}

The other system we analyzed were the companies of the S\&P500
consisting of $N=443$ companies. 
First
we looked at the case of strictly FM interactions, {\it i.e.} we took the
absolute value of the correlation coefficients. The
Fig.~\ref{fm_sp_khi} presents the $\hat{\chi}(T)$ function. The observed large
fluctuations in the susceptibility function is related to the narrow
distribution of correlation coefficients (Fig.~\ref{corr_distrib}).
This makes the separation of clusters sensitive to the temperature.

\begin{figure}
\centerline{
\epsfig{file=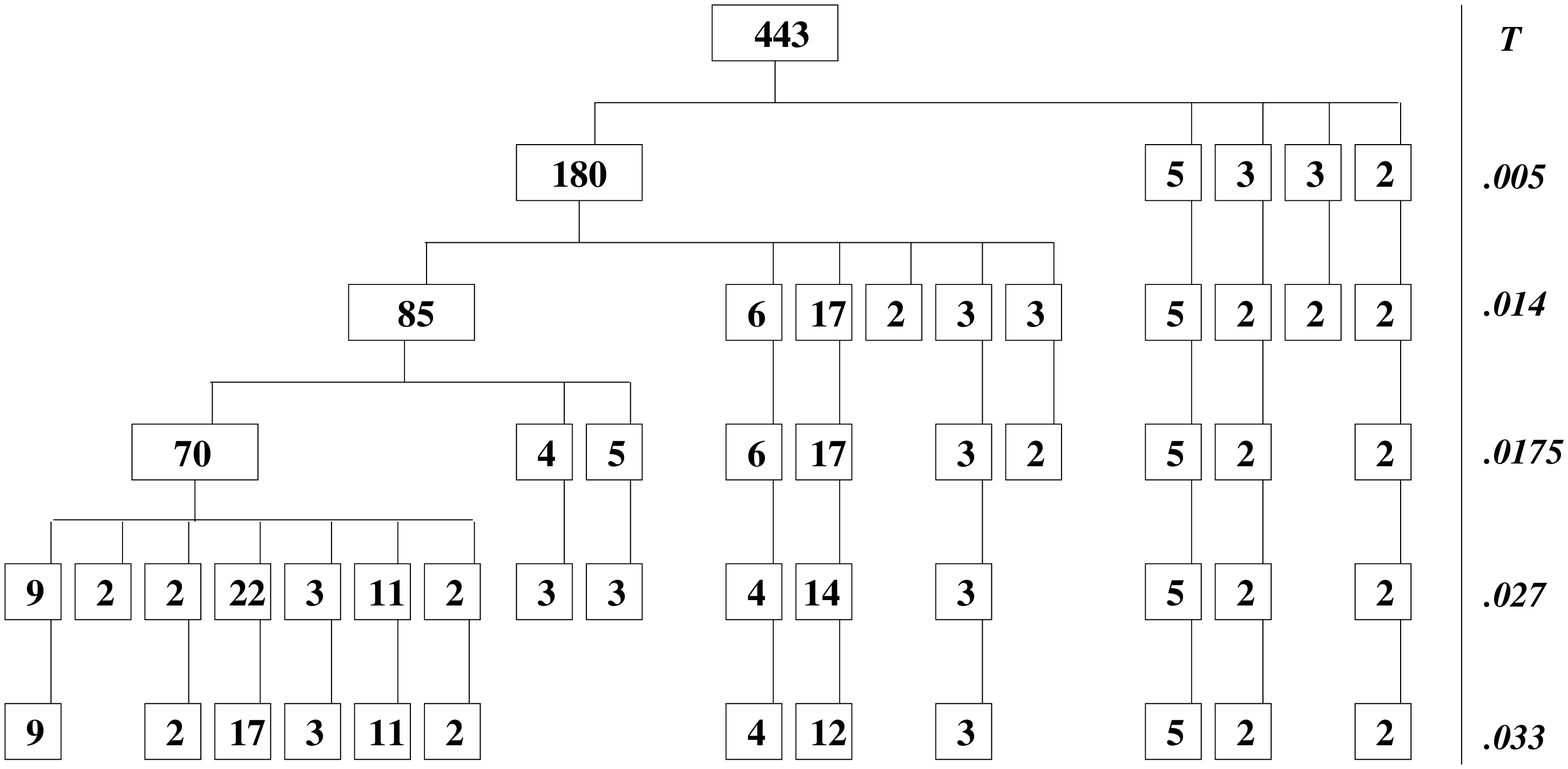,height=1.8truein,width=3.4truein}}
\centerline{}
\caption{\footnotesize  The hierarchical structure of clusters of the 
S\&P500 companies in the FM case. The clusters consisting of single 
companies are not indicated.}
\label{fm_sp_hier}
\end{figure}

This system represents a clear hierarchical structure of the
clusters, see Fig.~\ref{fm_sp_hier}. In the boxes the number of the
elements of the cluster are indicated. The efficiency of the method is
represented by the fact that the figure matches well with the
corresponding result obtained by the minimal spanning tree method, including
the specific composition of the clusters. {\it E.g.}, at the lowest level of
the hierarchy (highest temperature in the SPM phase) the different
industrial branches can be clearly identified: Oil, electricity, gold
mining, etc. companies build separate clusters.

It is misleading in the above picture that we take strong
anti-correlations equal to strong correlations: We have to take into
account the anti-correlations properly. To simulate an inhomogeneous Potts
model with long range and altering sign interactions seems to be a 
hard task, even the definition of the order parameter is non-trivial. 
Fortunately, in our case the negative interactions play a secondary
role as compared to the ferromagnetic ones: They are much
weaker and their number is much less.  This is shown in
Fig.~\ref{corr_distrib} where the distribution of the correlation
coefficients is presented and not the interactions.
Further simplification is that almost all the negative interaction is
associated with only five companies (gold mining) building one cluster
in the FM version too since there are quite strong positive
correlations between them.
The consequence is that except of very low temperatures there will be
no significant difference to the FM case and the simulation of our
system can be carried out by a simple Metropolis algorithm, there is
no need for the application of more sophisticated tools like the
multi-canonical algorithm.

However, the determination of the low temperature configurations is
not straightforward. The system falls easily into
a local minimum and it takes much simulation time to get out of
there. Therefore we used a process in the spirit of the simulated
tempering~\cite{KK}. We excite the system to a higher temperature level. Then
the temperature is lowered gradually so that at each temperature level
we keep the configuration according to the minimal energy and keep the
records of the best candidates of the low energy configurations.

\begin{figure}
\centerline{
\epsfig{file=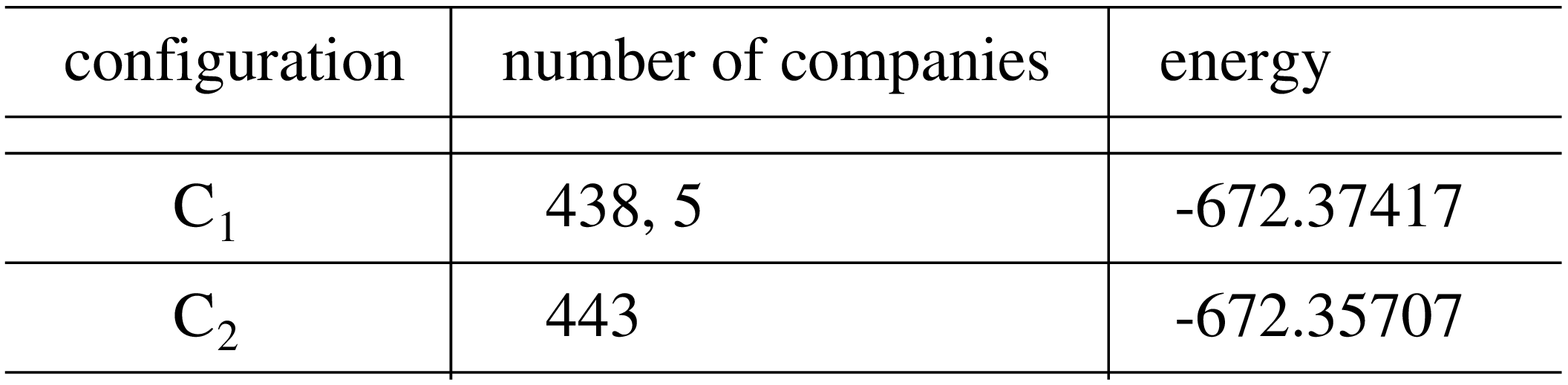,height=0.8truein,width=3.0truein}}
\centerline{}
\caption{\footnotesize Energies and configurations at low temperature 
in the AFM case of the S\&P500 companies.}
\label{energy_levels}
\end{figure}

Fig.~\ref{energy_levels} represents the configurations and their energy
values we got for this temperature range. 
Clearly, the pure FM
state ($C_2$) will not have the lowest energy value. This is not very
surprising if one knows that those five companies have to fall into
a separate cluster.

Our goal was to identify the clusters of companies at two stock
indices, and to show that the repulsion between the companies - due to
the negative correlation coefficients - can modify the cluster
structure.  Due to the distribution of the correlation coefficients in
our system this effect is significant only at the ground state,
{\it i.e.}, in the dominant cluster structure. Nevertheless, we think that
our analysis demonstrates the importance of the repulsion
effects in the clustering problem.

\section*{Acknowledgments}\indent
Partial support by OTKA-T029985 is acknowledged with thanks.

\end{multicols}
\end{document}